\def\simlt{\hbox{ \rlap{\raise 0.425ex\hbox{$<$}}\lower 0.65ex\hbox{$\sim$} }}
\def\simgt{\hbox{ \rlap{\raise 0.425ex\hbox{$>$}}\lower 0.65ex\hbox{$\sim$} }}
\def\that{{\hat t}}
\def\rs{U_*}
\def\msun{ {\rm M_\odot} }
\def\ee#1{\times 10^{#1}}
\def\etal{et~al.\ }

\documentstyle[aaspp4,epsf,astrobib,12pt]{article}
\begin{document}

\title{The MACHO Project: Limits on Planetary Mass Dark Matter
in the Galactic Halo from Gravitational Microlensing}

\author{
	  C.~Alcock\altaffilmark{1,2}, 
	R.A.~Allsman\altaffilmark{3},
	  D.~Alves\altaffilmark{1,4},
	T.S.~Axelrod\altaffilmark{5},
	  A.~Becker\altaffilmark{6},
	D.P.~Bennett\altaffilmark{1,2},
	K.H.~Cook\altaffilmark{1,2},
	K.C.~Freeman\altaffilmark{5},
	  K.~Griest\altaffilmark{2,7},
	  J.~Guern\altaffilmark{2,7},
	M.J.~Lehner\altaffilmark{2,7},
	S.L.~Marshall\altaffilmark{1,2},
	B.A.~Peterson\altaffilmark{5},
	M.R.~Pratt\altaffilmark{2,6,8},
	P.J.~Quinn\altaffilmark{5},
	A.W.~Rodgers\altaffilmark{5},
	C.W.~Stubbs\altaffilmark{2,5,6,8},
	  W.~Sutherland\altaffilmark{9}
	}
\begin{center}
{\bf (The MACHO Collaboration) }
\end{center}

\altaffiltext{1}{Lawrence Livermore National Laboratory, Livermore, CA 94550\\
	Email: {\tt alcock, alves, bennett, kcook, stuart@igpp.llnl.gov}}

\altaffiltext{2}{Center for Particle Astrophysics,
	University of California, Berkeley, CA 94720}

\altaffiltext{3}{Supercomputing Facility, Australian National University,
	Canberra, ACT 0200, Australia \\
	Email: {\tt robyn@macho.anu.edu.au}}

\altaffiltext{4}{Department of Physics, University of California,
	Davis, CA 95616 }

\altaffiltext{5}{Mt.~Stromlo and Siding Spring Observatories,
	Australian National University, Weston, ACT 2611, Australia\\
	Email: {\tt tsa, kcf, peterson, pjq, alex@merlin.anu.edu.au}}

\altaffiltext{6}{Departments of Astronomy and Physics,
	University of Washington, Seattle, WA 98195\\
	Email: {\tt becker, mrp, stubbs@astro.washington.edu}}

\altaffiltext{7}{Department of Physics, University of California,
	San Diego, CA 92039\\
	Email: {\tt kgriest, jguern, mlehner@ucsd.edu }}

\altaffiltext{8}{Department of Physics, University of California,
	Santa Barbara, CA 93106 }

\altaffiltext{9}{Department of Physics, University of Oxford,
	Oxford OX1 3RH, U.K.\\
	Email: {\tt w.sutherland@physics.ox.ac.uk}}

\begin{abstract} 
The MACHO project has been monitoring about ten million stars in the
Large Magellanic Cloud in the search for
gravitational microlensing events caused by massive compact
halo objects (Machos) in the halo of the Milky Way.
In our standard analysis, we have searched this data set for 
well sampled, long duration
microlensing lightcurves, detected several microlensing
events consistent with Machos in the
$0.1\,\msun \simlt m \simlt 1.0 \,\msun$ mass
range, and set limits on the abundance of objects with masses 
$10^{-5}\,\msun \simlt m \simlt 10^{-1} \,\msun$.
In this paper, we present a different type of analysis involving the
search for very short time scale brightenings of stars which is
used to set strong limits on the abundance of lower mass Machos.
Our analysis of the first two years of data toward the LMC indicates that
Machos with masses in the range
$2.5\ee{-7}\,\msun < m < 5.2\ee{-4} \,\msun$ cannot make up
the entire mass of a standard spherical dark halo.
Combining these results with those from the standard analysis,
we find that the halo dark matter may not be comprised of objects with
masses $2.5\ee{-7}\,\msun < m < 8.1\ee{-2}\,\msun$.
\end{abstract}

\section{Introduction}
\label{intro}

If a significant fraction of the dark halo of the Milky Way is made up
of Machos (MAssive Compact Halo Objects),
it should be possible to detect them by searching for
gravitational microlensing
\cite{paczynski86,petrou}. 
As a Macho passes near the line of sight
to a background star, the star appears to be magnified by a factor
\begin{equation}
A = {{u^2 + 2}\over{u\sqrt{u^2 + 4}}}
\label{eqamp}
\end{equation}
where $u=b/r_E$, $b$ is the distance from the Macho to the line of
sight, and the Einstein ring radius $r_E$ is given by
\begin{equation}
r_E = \sqrt{{4GmLx(1-x)}\over{c^2}}
\end{equation}
where $m$ is the mass of the Macho, $L$ is the observer-star distance, and
$x$ is the ratio of the observer-lens and observer-star distances.

Since Machos are in motion $(v\sim v_\odot = 220$km/sec) relative to
the line of sight, this magnification is time dependent, with
$A(t)=A(u(t))$, where
\begin{equation}
u(t)=\left[ u_{\rm min}^2 + \left({2 (t-t_0 ) \over \that}\right)^2
\right]^{1 \over 2}.
\label{equt}
\end{equation}
Here $t_0$ is the time of peak magnification, and $\that$
is the event duration, which can be
written
\begin{equation}
\that = 2r_E/v_{\perp} \sim 130\sqrt{m/\msun} \, \rm days,
\label{eqt_hat}
\end{equation}
where $v_{\perp}$ is the Macho velocity relative to the line-of-sight.
For more detailed information, see 
\citeN{paczynski86}
and
\citeN{griest91a}.

If the halo consisted entirely
of objects with masses under about $10^{-4} \,\msun$ the average duration
of microlensing would be less than 1.5  days,
and the events would last only about three hours if the halo were made
of $10^{-6} \,\msun$ objects. In order to clearly see the shape of the
microlensing curve for such low mass lenses, images of a lensed star
must be taken in rapid succession during the event. Such an
experiment was undertaken by the EROS collaboration
\cite{eros},
in which a total of about 82,000 stars were imaged up to 46 times per
night for several months. No microlensing events were found, and it
was reported that objects with masses
$5\times 10^{-8}\,\msun < m < 5\times 10^{-4} \,\msun$
can not comprise the entire Galactic Halo at the 90\% c.l.
We have not followed this approach, but we have used a different technique
also capable of setting limits on low mass Machos.

The MACHO collaboration has been monitoring the brightnesses of several million
stars in the LMC, SMC, and Galactic Bulge since 1992 July using the 50'
telescope at Mt. Stromlo, Australia. A dichroic beamsplitter and filters
are used to provide simultaneous measurements in red and blue
passbands \cite{lmc1}.
The observing strategy for the first
two years of LMC data was designed to be sensitive to objects with masses
$m > 10^{-3} \,\msun$, so a typical LMC field was generally imaged at most
once or twice per clear night. Therefore microlensing events with durations
under a few days will have very few magnified points on their light curves
and would show up in our data as upward excursions
of one, two or three consecutive measurements, occurring on stars which
otherwise appeared completely normal. 
In this paper we search specifically for such short duration ``spike" events.
Clearly, if any spikes are
detected, no conclusion could be drawn regarding their
origin since there would be insufficient detail in the
lightcurves. Therefore,
the technique described here is most useful when few if any
spikes are found, in which case useful upper limits can be
placed on the prevalence of low mass Machos. 
After applying selection criteria described below,
we do not find any such spikes
and so are able to strongly constrain the existence
of low mass objects in the halo of the Milky Way.

\section{Event Selection and Detection}
\label{selection}
The analysis reported here uses the first two years of LMC data
\cite{macho-prl95a,lmc2,lmc1}.
Twenty-two fields of 0.5 square degree each were monitored on
every clear night from 1992 July 20 to 1994 October 26,
for a total of 10827 observations.
A total of about 8.5 million stars are used in this analysis.  
The images are taken with a refurbished telescope system
\cite{telescope}
and a special purpose camera system 
\cite{macho-stubbs93,macho-marshall94},
photometrically reduced using
a special purpose code named Sodophot
\cite{bennett95},
and assembled
into time series for analysis.  
Each lightcurve consists 
of many measurements of the flux of a star in two filter bands (called
``red" and ``blue"), as well as estimated errors in the flux measurement
and several quantities used to detect probable systematic error in the 
measurement.  These quantities include the crowding, $\chi^2$ of PSF fit,
missing pixel fraction, cosmic ray flag, and sky background.
As in the standard analysis these measures are used
to remove suspect data before any further analysis is performed.

Also, as in the standard analysis, several properties of the expected
microlensing signal are used to eliminate
stars and events which are unlikely to arise from microlensing.
After imposing such selection criteria, it is necessary to calculate
the number of actual microlensing events which would have be removed
by these cuts, and this detection efficiency
calculation is discussed in the following sections.

Because this search is for short duration events,
one of the most powerful signatures of microlensing, the shape
of the lightcurve (eq's [\ref{eqamp}] and [\ref{equt}]), cannot be
used as a selection criterion.
Since there are three free parameters for the microlensing
lightcurve shape, we would need four or more observations during
the event to get a meaningful fit.
Therefore, any phenomenon which causes a significant upward excursion in one
or two observations could be mistaken for very short duration microlensing.  
Looking through our data, we
find many one-observation excursions.  A partial list of causes includes
satellite tracks and glints, telescope slips, and asteroids.
In order to reduce this background, 
we consider only those instances 
where two or three exposures were taken of the same same star on
the same night.  
Since each exposure gives  
both red and blue flux measurements, we
define a ``quad" as a sequence
of 2 (or 3) exposures of a star which are on the same night.
In the first two years of LMC data we have
$1.44 \times 10^8$ quads with two
measurements on the same night and
$5.8 \times 10^6$ quads with three.
We then require that all the measurements in the quad be
significantly magnified, and
we also make use of the fact that for very short duration
microlensing, the measurements of the star on the previous and following
nights should not have significant upward excursions.  
Thus each quad represents a potential detection of short
duration microlensing events,
and because the microlensing rate is proportional to $m^{-1/2}$
\cite{griest91a}
a substantial
number of quads would be magnified if the Milky Way Halo were made of
low mass Machos.

We define the magnification of a given measurement as
\begin{equation}
A = f / \bar{f},
\end{equation}
where $f$ is the flux of the measurement and $\bar{f}$ is the median flux
of all the points in the relevant pass band that pass the quality cuts
described above. 
In order to reduce the statistical probability
of random fluctuations in measured flux
giving false triggers, we require
that all four (or all six) measurements have positive excursions 
of more than $4\sigma$, where $\sigma$ is set by the flux measurement error.  
That is, we require
that the magnification be above a threshold magnification
$A_T$, with $A_T-1 = 4\sigma_{\rm max}$,
where $\sigma_{\rm max}$ is the largest magnification
error of the four or six quad measurements.
By using only these sets of
measurements and setting a threshold
proportional to the error,
the probability of false event detection can be greatly
reduced while still allowing a strong limit to be set. With a
threshold $A_T-1 = 4\sigma_{\rm max}$ and a total of $ 1.5\times10^8$ quads,
the expected number of
false triggers from statistical fluctuations is
$(3.2\times10^{-5})^4 \times 1.5\times10^8 = 1.6\ee{-10}$.
(This estimate assumes Gaussian errors, so realistically the
non-Gaussian tails will increase this number substantially.)
The $4\sigma_{\rm max}$ threshold was chosen {\em a priori} because after
investigation of several possible analysis methods on a small subset of
the data, it appeared most likely to give few (if any) events
and a significant detection efficiency. 

In addition to the above criteria,
we demand that there be a reasonable number of
high quality data points on the star in order to accurately
determine the baseline flux.
Thus we cut on the number of simultaneous red and blue data points and
the average photometric error.
In our data set we have found a large number of periodic
and non-periodic variables which can be eliminated due to the fact that
they vary continually, so we also demand that the lensed star not be
flagged as a variable.
Next, since microlensing is so rare, we do not expect more than one
microlensing event to take place on a given star, so we can also eliminate
stars in which two or more such events are found.
Finally, 
since gravitational lensing should magnify all wavebands by the same amount,
we can demand that the four or six spike points be achromactic within errors.
Therefore we make the following set of cuts:

1)  The star must have at least six observations in which both the red
and blue data points pass the cuts in crowding, seeing, etc.

2)  The star must have $\vr < 0.9$.

3)  Both the red and blue points in the measurements in the
    quad, immediately before the quad, and immediately after
    the quad  
    must pass the cuts in crowding, seeing, etc.
    and have a magnification error less than 0.5.

4)  The ``robust" $\chi^2$ (that is, a $\chi^2$ fit with the highest
    and lowest ten percent of the data excluded)
    of a fit to a constant flux must be less than 0.9 in
    both red and blue.
    
5)  $A-1 > 4\sigma_{\rm max}$ for all points in the quad,
where $\sigma_{\rm max}$ 
    is the maximum error of the measurements in the quad.

6)  $A-1 < 4\sigma_{\rm max}$ for the red and blue points in
    the measurement
    previous to the quad and the measurement
    following the quad.

7)  the measurements in the quad must be achromatic within errors, that is
    $\Delta < 2\sigma_\Delta$, where $\Delta = |A_r /A_b - 1|$.

8)  There may be at most one event per star.

\begin{figure}
\plotone{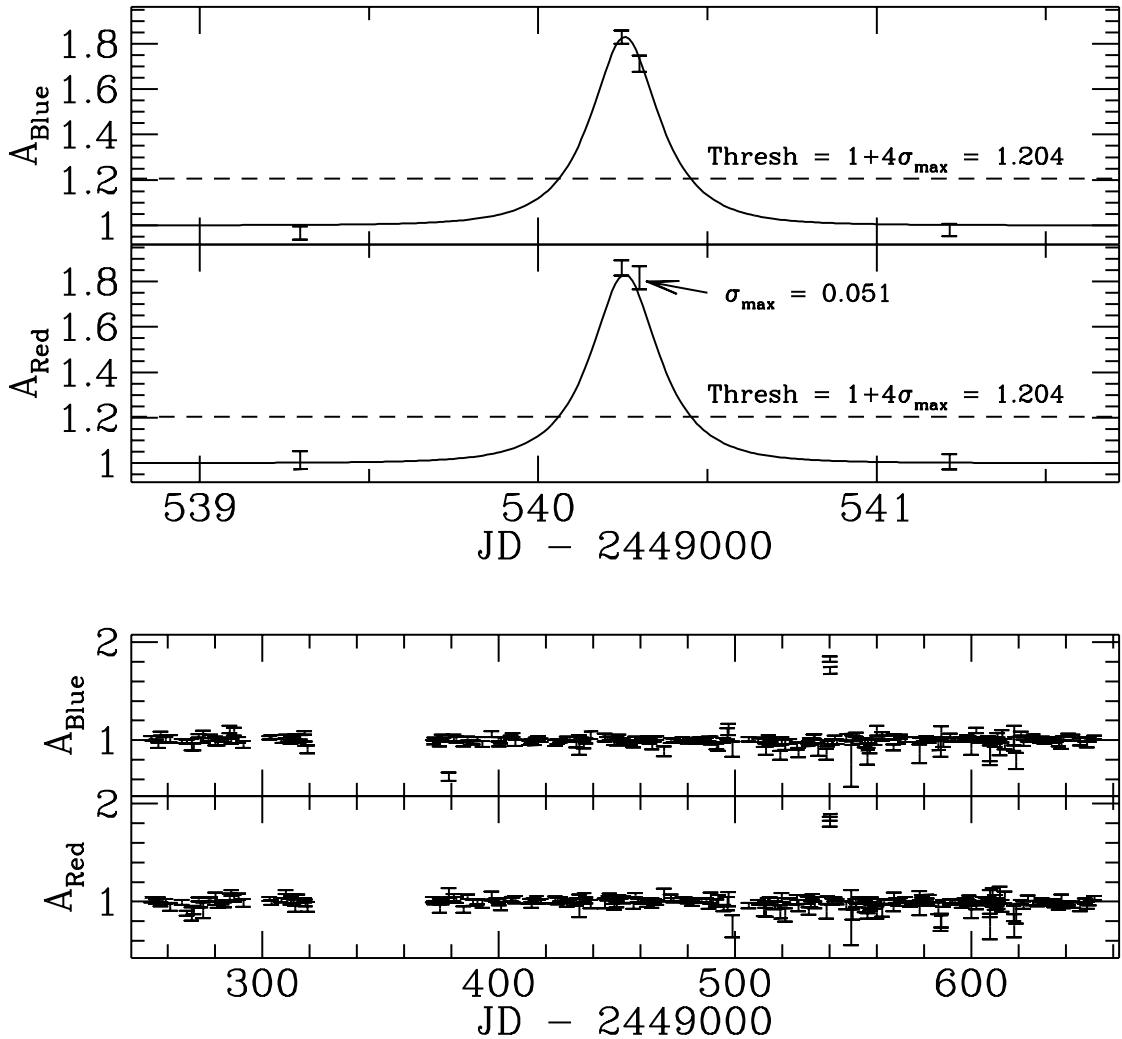}
\caption{
A typical Monte Carlo event which
passes the
cuts used in this analysis.
The top two curves
are an expanded view of the quad in which the event occurred. Note that
the points in the quad are above the threshold of 1.204, and the
previous and following measurements are below it. The solid line is
the theoretical microlensing curve added to the data. On the bottom
are the entire year 2 lightcurves for the same event. The magnified
quad points are clearly visible at day 540.
}
\label{figdat}
\end{figure}

Cuts 1, 2, 4, and 8 are for the elimination of false events caused by variable
stars, cuts 1 and 3 are to eliminate background
caused by poor photometry, cut 5 is to
reduce the chance of statistical fluctuations 
or single-observation glitches causing an event, and cut 6
is to eliminate longer duration events. See Figure \ref{figdat} for an
example of our data and a Monte Carlo event which passes these cuts. 

These cuts were run on the first two years of LMC data, and no events were
found.
To test the robustness of these cuts, the analysis was run with thresholds
varying between $3\sigma_{\rm max}$ and
$5\sigma_{\rm max}$. While a few events were found
at thresholds below $4\sigma_{\rm max}$, none were found at the {\em a priori} 
threshold of $4\sigma_{\rm max}$. This will be discussed further in
Section \ref{results}.

\section{Theoretical Event Rates}
\label{theory}

In order to use our non-detection of spike events to make a statement about 
the content of the Milky Way halo, we need to predict the number
of events we would have expected to find if the halo is made
of low mass Machos.  
Thus we need to know
the efficiency with which our experiment, in combination with the above
selection criteria, would have detected short duration microlensing events
if in fact such events occurred.
The search for spike events is sensitive only to durations 
of about 0.1 to 4 days, corresponding to masses of about $10^{-6}$ to
$10^{-3} \, \rm M_\odot$ (see eq. [\ref{eqt_hat}]),
and we need to use a halo model to make a connection
between event duration, rate, and Macho mass.  

We first consider a simple spherical halo model with mass density
\begin{equation}
\rho(r)= \rho_0 {{R_0^2 + a^2} \over {r^2 + a^2}}
\label{eqmodel}
\end{equation}
where $\rho_0 = 0.008\,\msun\rm pc^{-3}$ is the local dark matter mass density,
$r$ is the distance to the center of the Galaxy,
$R_0 = 8.5\,\rm kpc$
is the distance of the sun from the Galactic center, 
and $a = 5\,\rm kpc$ is the Galactic core radius.
\citeN{griest91a}
showed that the microlensing rate of a $\delta$-function mass
distribution is given by
\begin{equation}
\Gamma = 1.60\times 10^{-6} u_T/\sqrt{m/\msun} \,\rm events/year,
\end{equation}
where $u_T = u(A_T)$, and $A_T$ is the magnification threshold for
an event.
(The expression for $\Gamma$ is slightly different from that given in
Griest (1991) because we use $50$ kpc for the distance to the LMC, rather
than $55$ kpc.)
In an experiment the number of microlensing events which are expected
to be detected is given by
\begin{equation}
N_{\rm exp} = \Gamma E {\cal E}
\end{equation}
where $\Gamma$ is the microlensing rate (in event/year/star), 
$E$ is the effective exposure (in star-years)
and ${\cal E}$ is the average detection efficiency.

The probability of an event occurring with duration
$\that_{\rm min} < \that < \that_{\rm max}$
can be written
\begin{equation}
P = {1\over\Gamma} \int\limits_{\that_{\rm min}}^{\that_{\rm max}}
{{d\Gamma}\over{d\that}}
   d\that.
\end{equation}
where, in the approximation of a stationary line of sight, 
the distribution of event durations is given by
\cite{lmc1}

\begin{equation}
{d\Gamma \over d\that} = {{32 u_T L \rho_0} \over {m v_0^2 \that^4}}
\int\limits_0^1 dx {r_E^4(x)Ae^{-Q} \over A + Bx + x^2}
\label{eqrate}
\end{equation}
where 
$A = (R_0^2 + a^2)/L^2$, $B = -2r_0\cos b \cos l / L$,
$b$ and $l$ are the galactic coordinates of the source star,
$v_0 = 220\,\rm km/sec$ is the solar circular velocity,
and $Q = 4r_E^2(x)/(v_0^2\that^2)$.
The probability that such an event is actually observed depends
strongly on the duration, since the spike selection criteria above
eliminate the possibility of detecting long duration events or events which
last only an hour. We can thus define the average efficiency of observing
such events as
\begin{equation}
{\cal E} = {1 \over \Gamma} \int\limits_{\that_{\rm min}}^{\that_{\rm max}}
{d\Gamma \over d\that} \epsilon(\that) d\that
\label{eqprob}
\end{equation}
where $\epsilon(\that)$ is the detection efficiency as a function
of event duration.

\section{Finite Source Effects}
\label{fs}

When the impact parameter of the lens is comparable to the size of the
lensed object, the magnification can differ significantly from the point
source approximation given in equation (\ref{eqamp}). For a lensed star with
radius $R_*$, we define
\begin{equation}
\rs = {R_*x \over r_E(x)}
\end{equation}
as the ``effective radius'' of the star (the radius of the star normalized
to the Einstein radius and scaled to the lens plane). The point
source approximation will then break down completely for $u \simlt \rs$
because only the fraction of the surface of the star inside the Einstein
ring radius will be significantly magnified.
In the case of a star of constant surface brightness, we use
\cite{witt94}
\begin{eqnarray}
A & =&  {2 \over \pi \rs} + {1 + \rs^2 \over \rs^2}\left({\pi \over 2}
+ \arcsin {\rs^2 - 1 \over \rs^2 + 1}\right)
\end{eqnarray}
for $u = \rs$, and 
\begin{eqnarray}
A & = & {2(u-\rs)^2 \over \pi \rs^2(u+\rs)}
{1+\rs^2 \over \sqrt{4 + (u - \rs)^2}} \ \ \Pi \left( {\pi \over 2}, n, k\right)
\nonumber \\
&   &  + {u+\rs \over 2\pi \rs^2}\sqrt{4 + (u-\rs)^2}
\ \ E\left({\pi \over 2}, k\right) \nonumber \\
&   & - {u-\rs \over 2\pi \rs^2}\;{8+(u^2 - \rs^2) \over \sqrt{4+(u-\rs)^2}}
\ \ F\left({\pi \over 2}, k\right)
\end{eqnarray}
for $u \neq \rs$, where 
\begin{eqnarray}
n & = & {4u\rs \over (u + \rs)^2}, \nonumber \\
k & = & \sqrt{ {4n \over 4 + (u - \rs)^2} }, \nonumber
\end{eqnarray}
and $F$, $E$, and $\Pi$ are elliptic integrals of the
first, second, and third kind.

\begin{figure}
\plotone{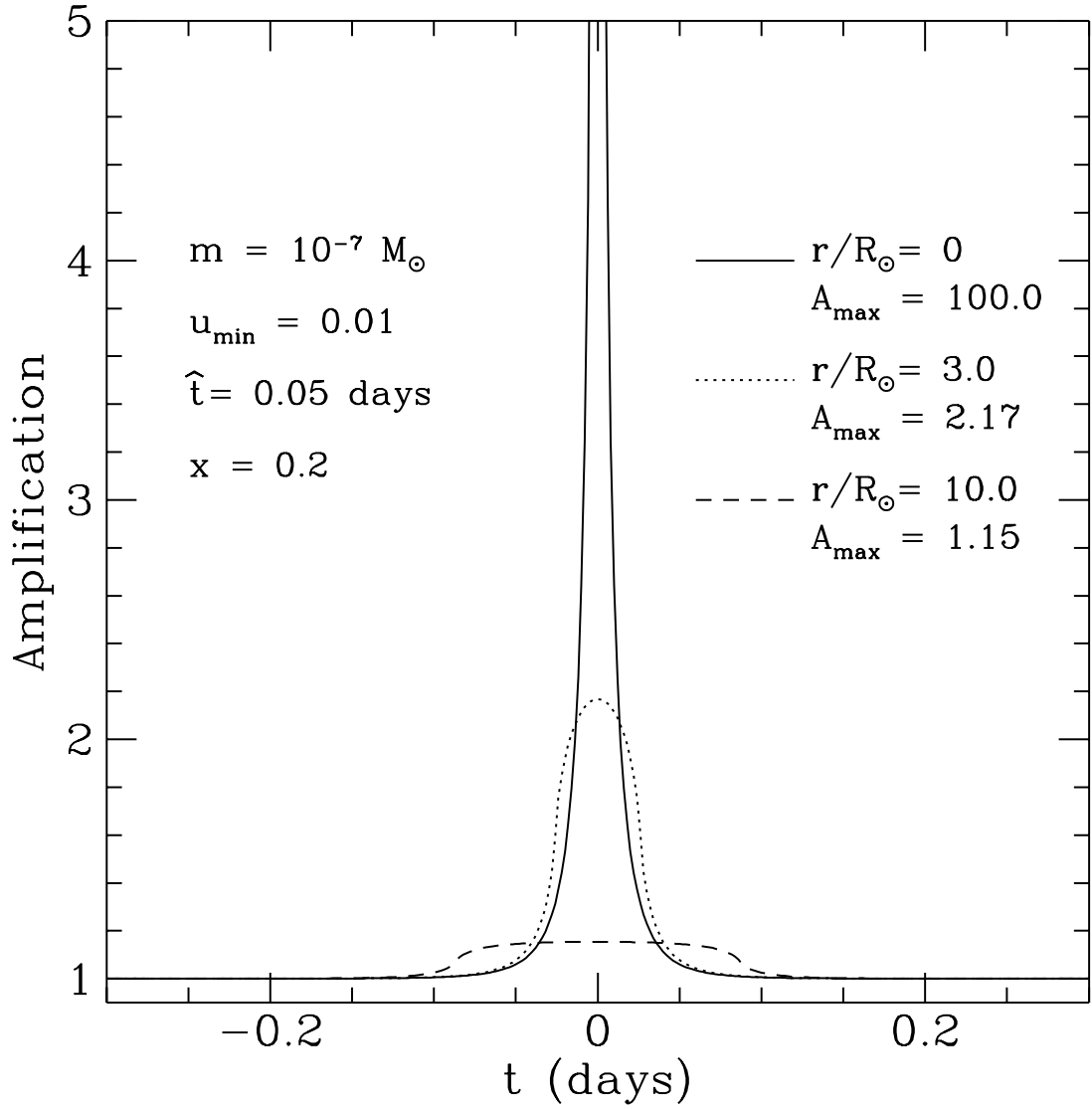}
\caption{A plot of light curves for various source radii.
For larger sources, the maximum magnification decreases, but the width
of the curve increases.}
\label{figfs}
\end{figure}

There are two effects from the finite source size. First, the maximum
possible magnification
\begin{equation}
A_{\rm max} = {\sqrt{4 + \rs^2} \over \rs}
\end{equation}
becomes very low for lower mass lenses. For a star with $R = 10\, \rm R_\odot$
and a lens at $x = 0.5$, we have $A_{\rm max} = 18.3$ for a lens with
$m = 10^{-4} \,\rm M_\odot$. However, with $m = 10^{-6} \,\rm M_\odot$ we have
$A_{\rm max} = 2.08$ and with $m = 10^{-7} \,\rm M_\odot$ we get
$A_{\rm max} = 1.15$. Because we are searching for significant magnifications,
this effect would tend to lower the detection efficiency for lower mass
lenses. The second effect of a finite source size is that the star is
magnified for a longer period of time. This occurs because a fraction
of the star can be close enough to the lens to be significantly magnified
even if the lens is far from the center of the star (see Figure \ref{figfs}).
This effect increases the detection efficiency for low mass lenses whose
average event duration is shorter than the minimum detectable point
source event time scale. These two effects
mostly cancel each other out, with the detection efficiency increasing
slightly for very low mass Machos.

It is clear that the shape of the light curve is strongly dependent
on the lens distance $x$, so the detection efficiency is a function
of both $x$ and $\that$. Since $\that$ is a function of both $x$ and
$v_\perp$, we use $\epsilon = \epsilon(x, v_\perp)$ in order to
simplify the efficiency analysis. The average detection
efficiency (eq. [\ref{eqprob}]) then becomes
\begin{equation}
{\cal E} = {1 \over \Gamma} \int\limits_0^1 dx
\int\limits_0^{v_{\rm max}} dv_\perp 
{d\Gamma \over dx\,dv_\perp} \epsilon(x, v_\perp)
\label{eqeffxv}
\end{equation}
where $v_{\rm max}$ is some upper limit on the perpendicular
velocity of the lens.
The differential rate used in equation (\ref{eqeffxv}) can be found
from equation (\ref{eqrate}) using a simple change of variables:
\begin{equation}
{d\Gamma \over dxdv_{\perp}} = {{2 u_T L \rho_0 v_\perp^3 \that}
\over {m v_0^2}}
{A e^{-v_\perp^2 / v_0^2} \over A + Bx + x^2}.
\label{eqratexv}
\end{equation}

\section{Monte Carlo and Detection Efficiency}
\label{efficiency}

To measure the detection efficiency $\epsilon( x, v_\perp )$, 
a Monte Carlo
simulation was performed in which randomly generated
microlensing events were added to each star in the database. 
Then the same analysis used to search for spike events was performed
on these simulated data sets.
As a function of the lens position and perpendicular velocity of
these events one can then find
the fraction of simulated events which were recovered and define this
as $\epsilon(x, v_\perp)$.
From simple geometry one expects microlensing events to have a uniform
distribution in minimum impact parameter $u_{\rm min}$. A minimum error
of $0.014A$ is added to each data point, so the minimum
$3\sigma_{\rm max}$ threshold is given by $A_T = 1.042$, or
$u_{\rm min} = 2.262$.
(As stated previously, the analysis was run on several
thresholds varying between
$3\sigma_{\rm max}$ and $5\sigma_{\rm max}$, so the minimum threshold was used
to set the upper limit of the $u_{\rm min}$ distribution. This somewhat lowers
the detection efficiency for higher thresholds, but the differential
microlensing rate given in eq. [\ref{eqratexv}] is correspondingly higher
due to the factor of $u_T$ and there is no net effect in the final result.)
Therefore, in performing the Monte Carlo, the simulated
microlensing events were added with a uniform
distribution of $u_{\rm min}$ from 0 to 2.262.
To adequately sample the widest possible range of event durations
and finite source lightcurve shapes, the 
events were generated using
a distribution of $x$ which was uniform over 
$0 < x < 1$, and a $v_\perp$ distribution uniform over
$0 < v_\perp < v_{\rm max} = 667 \,\rm km/sec$.
In order to improve statistics the simulated events were
forced to peak during a quad
(that is,
$t_0$ between the time of measurements previous to and following a quad).
The total exposure time used to calculate $N_{\rm exp}$ was adjusted accordingly
by using the total ``quad time'' rather than the length of the observing
run.  Thus no simulated events were added during weeks when the telescope
was down.

The shape of the light curve is a function of the radius of the
star for low mass Machos, and the radius of a star is correlated with
its magnitude. Because brighter stars tend to give lower errors in
measured magnification, it follows that the shape of a light curve is
correlated with the event detection threshold. Therefore
the radius of the source
was estimated from its color and magnitude,
and this radius was used in conjunction with the observer-lens
distance $x$ to determine the shape of the simulated light curves.
Although limb darkening can change the shape of the light curve
\cite{witt94},
limb darkening coefficients are not well known for such a large
sample of stars. An investigation into the effect of limb darkening
has shown that the resulting uncertainties in magnification
are much smaller than those caused by the uncertainty in the radius
of the source star, so the effects of limb darkening are ignored
in this analysis.

Since the detection threshold is proportional to the maximum error of the
points in a quad, it is important to treat the errors correctly when
adding a fake microlensing event to a light curve. The error of the
magnification $A$ can be approximated by

\begin{equation}
  \sigma = (\sigma_s^2 + (0.014f/\bar{f})^2 + f/\bar{f}^2)^{1\over 2}
\end{equation}
where $\sigma_s$ is the error from sky background,
$f$ is the total
measured flux of the star, and $0.014f/\bar{f}$ is the minimum error added 
after photometric analysis.
The minimum error is also added in the standard analysis photometry
\cite{macho-nature93,lmc1,macho-bulge1}
in order to account for several
sources of systematic error in the photometry.
There are two limiting cases of the above formula: ``sky-dominated"
error and ``flux-dominated" error. In the case where the error comes mostly
from the sky subtraction, the error in the flux
does not change significantly when flux is added to create a simulated
microlensing event. 
On the other hand, when the
error is dominated by the Poisson statistics of the flux, the
error in the added flux should be scaled by $\sqrt{f}\,$.
Because the flux error is larger, giving higher thresholds,
this case was used in the Monte Carlo 
to get a conservative measure of the efficiency.
(The minimum error was subtracted before the
errors were adjusted and then added in again afterward, 
just as the minimum error 
is added in after the photometric reductions in the standard analysis.) 
The true efficiency
is expected to be closer to the flux error case, because in
general the sky error contributes significantly only to dim stars
which already have a fairly low efficiency due to
their large error bars (large threshold).

Because of our crowded fields and poor seeing, many of our photometric
objects are actually blends of two or more stars. When a blended
star is lensed, the measured magnification can be significantly smaller
than the true magnification. To quantify this effect, artificial
stars were added to real images taken under a variety of observing
conditions and the photometry code was run again. We thus created a
series of response functions of recovered vs. added flux, and when a
simulated microlensing event is generated the photometric object is
matched to one of these response functions using the object's magnitude.
The observing conditions on each point in the light curve are then
matched to similar conditions in the response function, and the recovered
flux is added to the data. A detailed description of this analysis
can be found in
\citeN{lmc1}.

\begin{figure}
\plotone{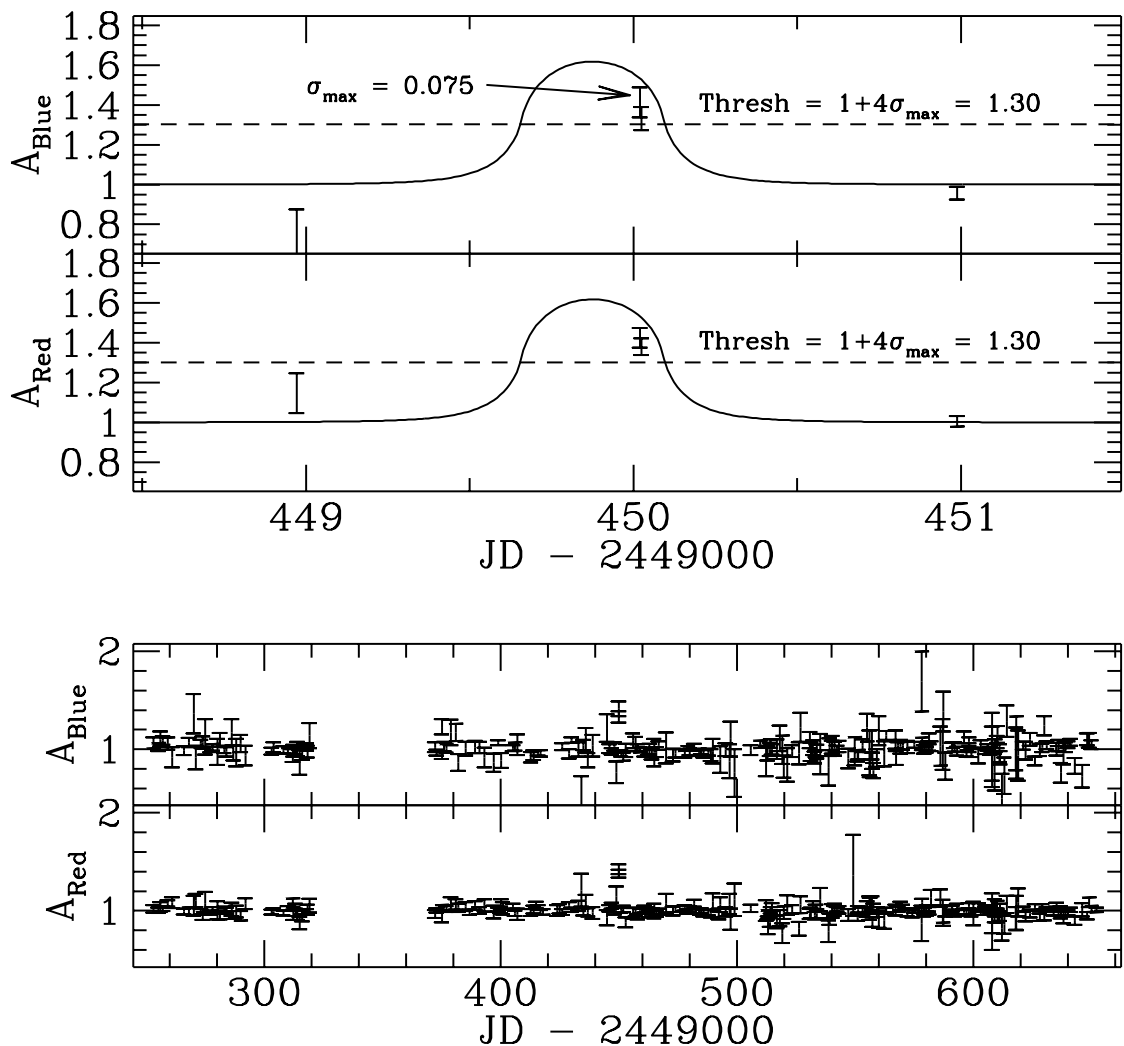}
\caption{A Monte Carlo event with significant blending
and finite source effects. The shape of the theoretical light curve
differs from a simple point source light curve, and the magnified
points are below the added magnification because only a fraction of
the object is lensed. The event still passes the cuts, and the
magnified quad can be seen at day 450 on the entire year 2
light curve.}
\label{figdatfs}
\end{figure}

The effects of blending on the detection efficiency are twofold.
First, the lower measured magnification will lower the efficiency
as fewer quads will be above the threshold magnification. However,
if an object is a blend of two or more stars, then there are two or
more stars that may be lensed. Thus the total number of stars in the data set
is larger than the number of photometric objects, and our total
exposure is significantly increased. A Monte Carlo event with a large
blend fraction (and significant finite source effects) can be seen in
Figure \ref{figdatfs}.

\section{Results} 
\label{results}

Because the blending effects described above are a function of the magnitude
of the lensed object, it follows that the detection efficiency depends
on the magnitude of the source as well as the lens position and velocity.
Using an infinitesmal bin width,
the efficiency $\epsilon (x, v_\perp, M)$ can be written
\begin{equation}
\epsilon = {dN_{\rm rec}/dx dv_\perp dM \over dN_{\rm add}/dx dv_\perp dM}
\end{equation}
where $M$ is the average of the red and blue magnitudes of the
lensed star (this value is the unblended stellar magnitude which is
determined by the blending
response function),
$N_{\rm add}$ is the number of fake events added to the data,
and $N_{\rm rec}$ is the number of these events recovered by the analysis.
To calculate the number of expected events, one must integrate
the efficiency over the stellar luminosity function $n(M)$:
\begin{equation}
N_{\rm exp} = \int dM \int\limits_0^1 dx
\int\limits_0^{v_{\rm max}} dv_\perp
{d\Gamma \over dx dv_\perp}
\epsilon(x, v_\perp, M) n(M) T
\end{equation}
where $T$ is the effective
exposure, or ``quad time'' of a given star.
To calculate the efficiency, we have
\begin{equation}
{dN_{\rm rec} \over dx dv_\perp dM} = \sum_i \delta(x-x_i)\delta(v_\perp -
v_{\perp i})\delta(M - M_i)
\end{equation}
where the sum is over all recovered events.
Because we add one simulated event to each object in the database,
we can write
\begin{equation}
{dN_{\rm add} \over dM} = n_{\rm sod}(M)
\end{equation}
where $n_{\rm sod}(M)$ is the Sodophot object luminosity function
(that is, the luminosity distribution of objects recovered by the
photometric reductions of the images).
This function is used because the response function stellar magnitudes
follow the Sodophot object distribution, and the response functions are
chosen uniformly from this distribution
\cite{lmc1}.
The events are added uniformly in $x$ and $v_\perp$, 
which gives
\begin{eqnarray}
{dN_{\rm add} \over dx dv_\perp dM} & = & {d^2 \over dx dv_\perp}
n_{\rm sod}(M) \nonumber \\
& = & {\rm const.} \times n_{\rm sod}(M) \nonumber \\
& = & {n_{\rm sod}(M) \over v_{\rm max}}
\end{eqnarray}
where the factor $1/v_{\rm max}$ is a normalization constant.
The number of expected events then becomes
\begin{equation}
N_{\rm exp} = \sum_i\left.{d\Gamma \over dx dv_\perp}\right|_{x_i, v_{\perp i}}
{n(M_i) \over n_{\rm sod}(M_i)} T_i v_{\rm max}
\label{eqnexp}
\end{equation}

Approximately 6.5\% of the stars used in this analysis can be found in more
than one field (Alcock \etal 1996b),
and we must scale the number of expected events accordingly.
However, the possibility of double counting in this analysis occurs only when
double or triple exposures are taken on overlapping fields on the same night.
This only happens in about $2/3$ of our quads, so we subtract 4.4\% from
our number of expected events rather than 6.5\%.

\begin{figure}
\plotone{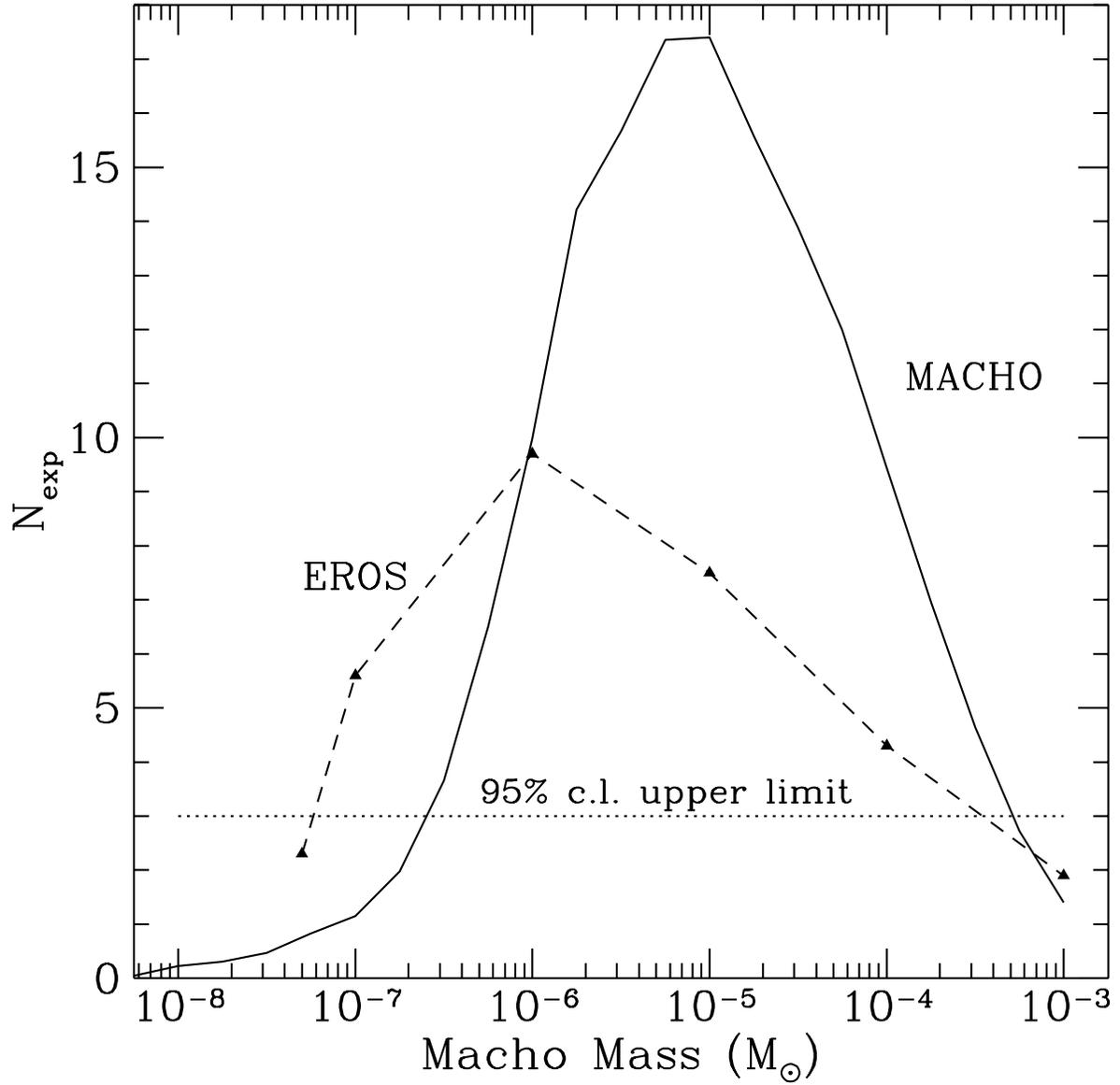}
\caption{A plot of the number of expected events vs mass
for a $\delta$-function mass distribution. With no events found the
95 \% c.l. upper limit is 3 events, and the region of the curve
above this limit are excluded. Also shown is the number of expected events
from the EROS CCD experiment.}
\label{fignexp}
\end{figure}

\begin{figure}
\plotone{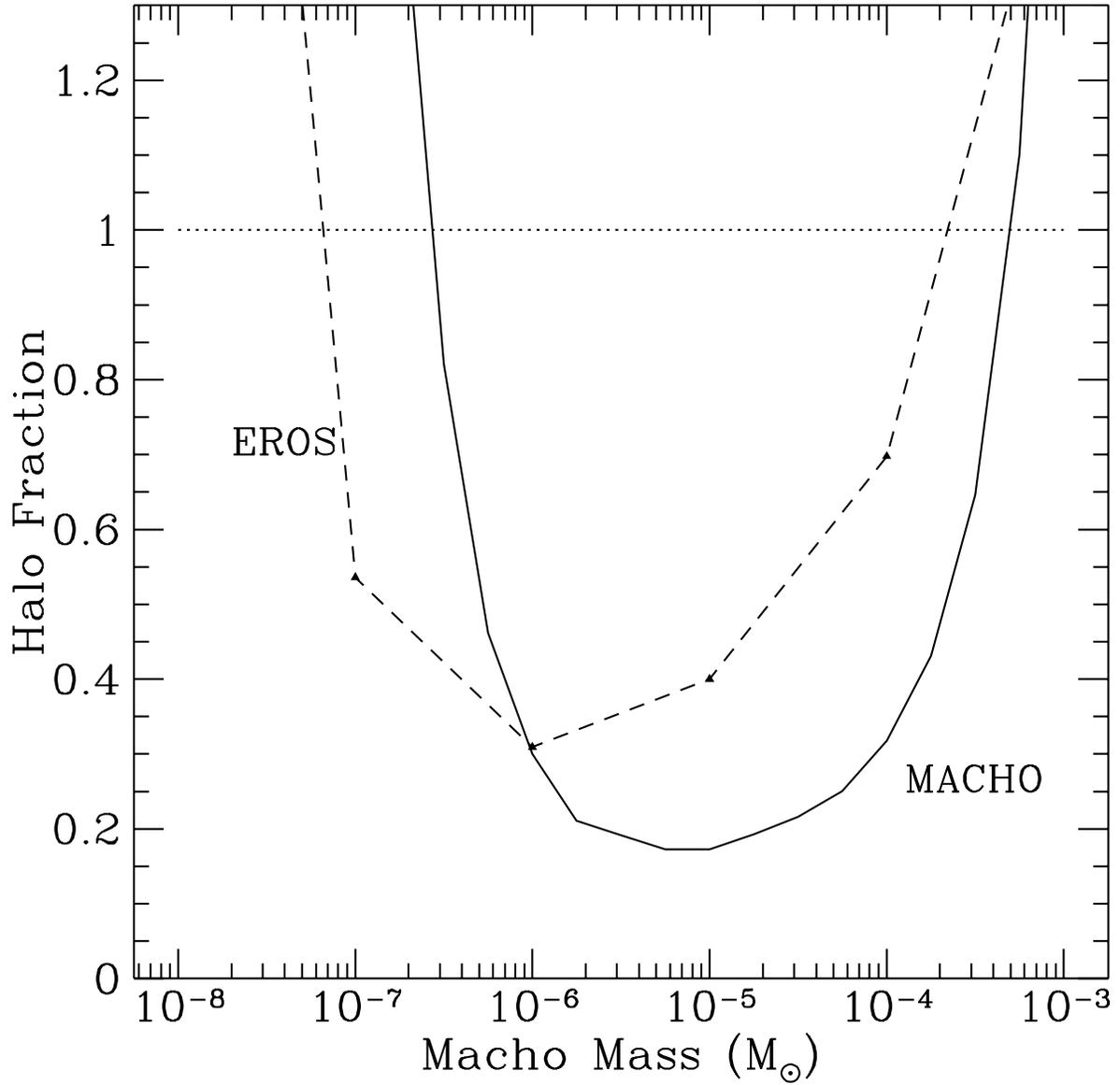}
\caption{A plot of allowed halo mass fraction vs mass
for a $\delta$-function mass distribution. The region above the solid line
is excluded at the 95\% c.l. Also shown is the halo fraction upper limit
from the EROS CCD experiment.}
\label{figfrac}
\end{figure}

A plot of $N_{\rm exp}$ vs mass 
for a $\delta$-function mass distribution is given in Figure \ref{fignexp}.  
At the peak at $10^{-5} \rm M_\odot$,
about 17 events would be expected to have been found if the halo is as
modeled in equation (\ref{eqmodel}),
and consisted entirely of Machos of that mass.   
Equivalently, we may convert the number of expected events into upper limits
on the allowed halo mass
fraction that can be contributed from objects in the
excluded mass range. Such a plot is given in Figure \ref{figfrac}.
Using the fact that no events were found we  
can place strong limits on the halo of the Milky Way.
The 95\% Poisson c.l.
when $N_{\rm obs}=0$ is $N_{\rm exp}= 3$ events,
so for the simple spherical halo model
masses between $2.5\ee{-7}\,\msun$
and $5.2\ee{-4}\,\msun$ are ruled out
at the 95\% c.l.
Although these limits are for a $\delta$-function mass distribution,
any model distribution containing a combination of masses in this range
is also ruled out at the 95\% c.l.
\cite{griest91a,lmc1}.

Also shown in Figures \ref{fignexp} and \ref{figfrac} are
the results from the EROS
CCD experiment
\cite{eros}.
Although they give a stronger limit for
$m < 10^{-6} \,\rm M_\odot$, the limits set by this analysis
give the strongest limits to date for $10^{-6}\,\msun \simlt m \simlt
10^{-3} \,\msun$.

\begin{figure}
\plotone{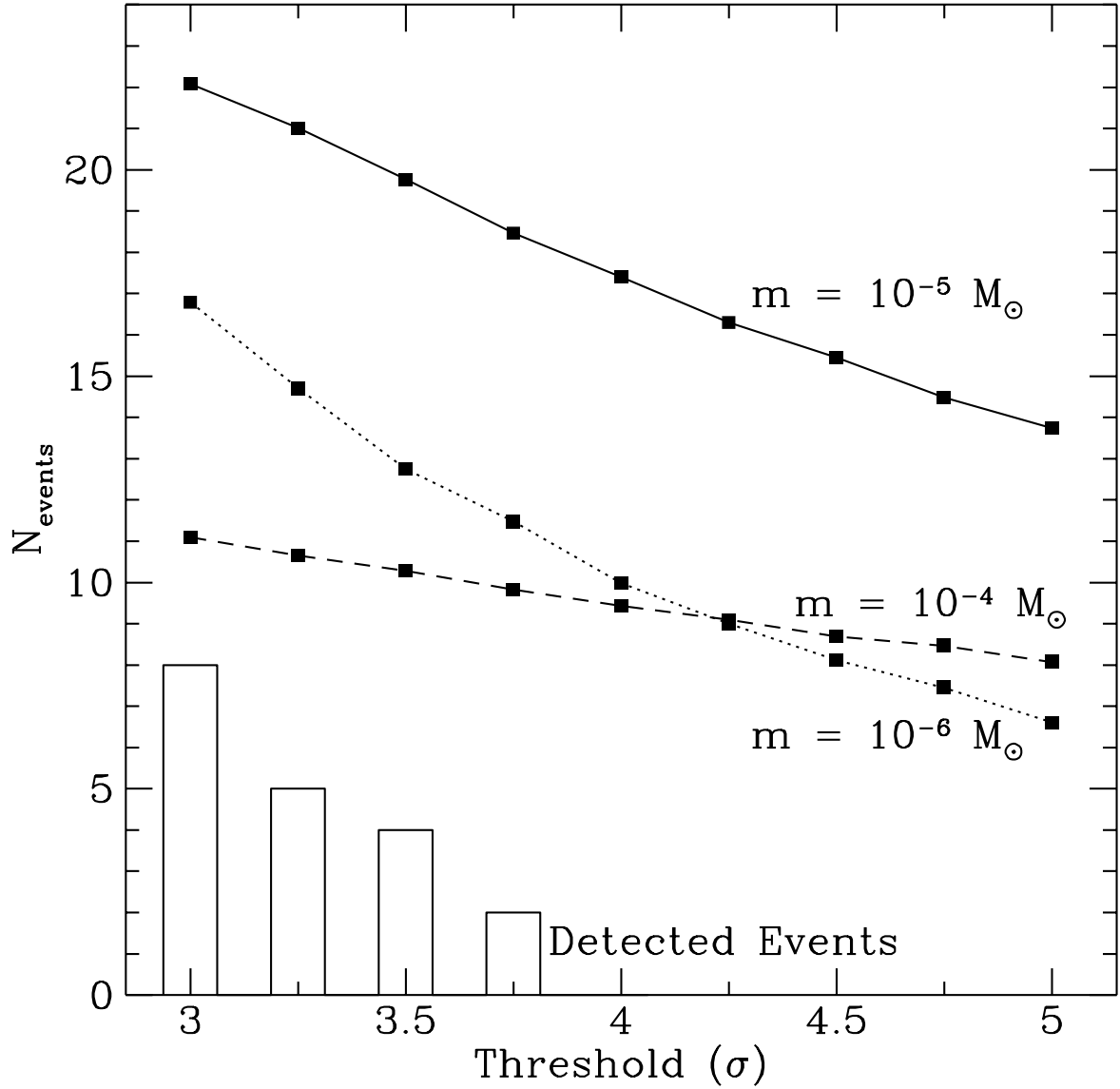}
\caption{Number of expected events as a function of threshold
for three values of Macho mass. Also shown are the number of observed events
as a function of threshold.}
\label{fignvst}
\end{figure}

\begin{figure}
\plotone{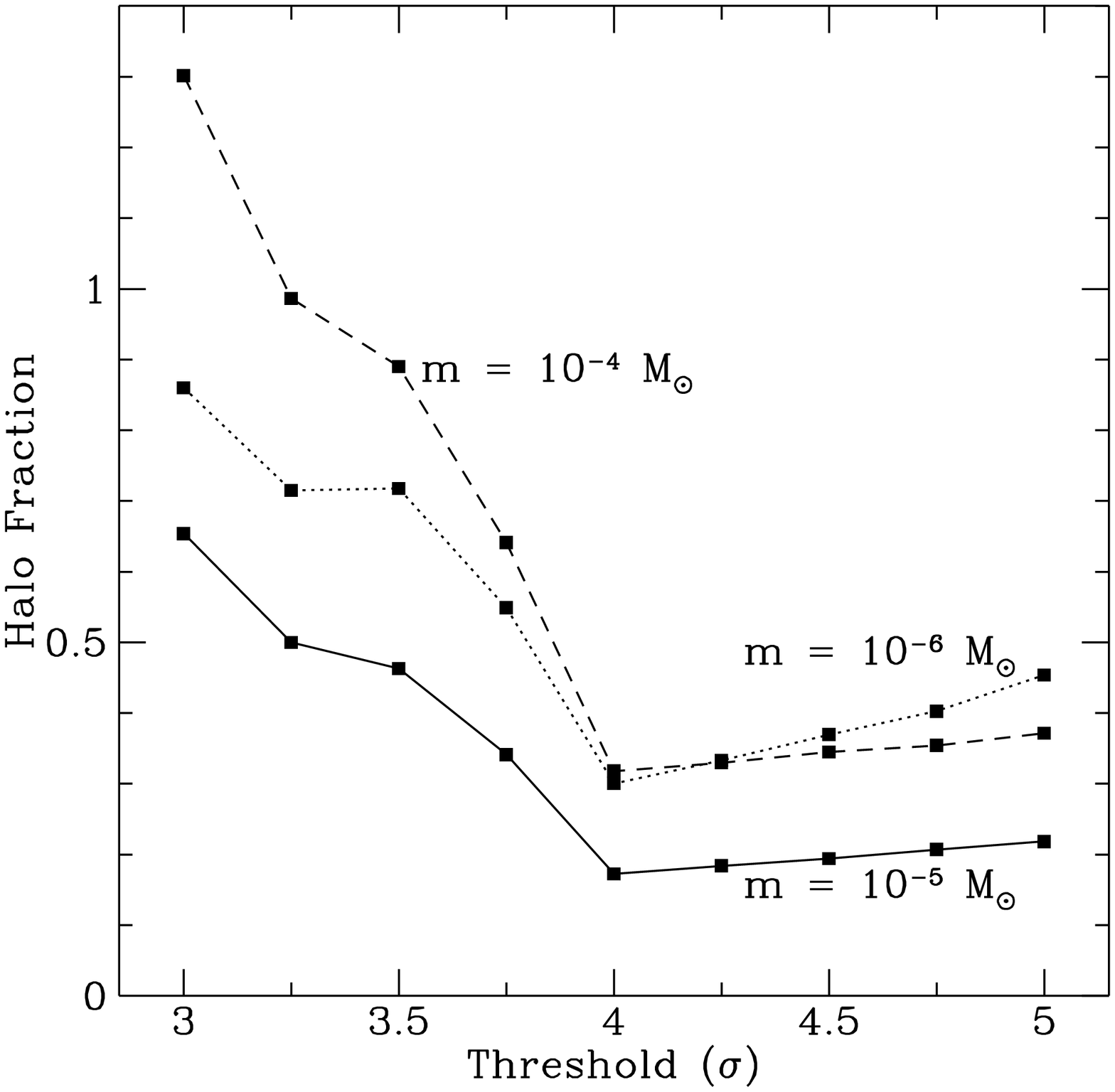}
\caption{Halo fraction upper limit (95\% c.l.)
as a function of threshold for three values of lens mass. The upper limit
rises at lower thresholds because events are detected at these thresholds
and the 95\% c.l. upper limit on $N_{\rm exp}$ increases accordingly.}
\label{figfvst}
\end{figure}

As mentioned previously, the analysis was run with thresholds varying
from $3\sigma_{\rm max}$ to $5\sigma_{\rm max}$ in order to determine
the robustness of the analysis. A plot of the number of expected and observed
events as a function of threshold can be found in Figure \ref{fignvst},
and the corresponding plot of halo fraction upper limit vs. threshold
can be found in Figure \ref{figfvst}. A total of 11 events were found
which passed various thresholds between $3\sigma_{\rm max}$ and
$3.75\sigma_{\rm max}$, but no events were found at thresholds of
$4\sigma_{\rm max}$ and higher. Of the 11 events found, 8 were on stars
with $V < 17.5$ and are likely low level variables which fell through
the variable star cuts. Of the three remaining events, two occurred in the
same field on the same night which indicates possible problems with
the observations, and inspection of the second image in the quad shows a
likely telescope slip during the exposure.
The remaining event passes only the $3\sigma_{\rm max}$ threshold cut.
In order to reduce these backgrounds in
future analysis runs, stars with $V < 17.5$ will be cut as will
images with more than one event, and it has been determined that these cuts
will reduce the number of expected events by about 20\%. However,
because neither the number of expected events or the number of observed
events varies drastically with threshold, the analysis using
the {\em a priori} $4\sigma_{\rm max}$
threshold is robust and the limits set on the abundance of
low mass Machos are valid.

\section{Combined Analyses}
\label{combanal}

\begin{figure}
\plotone{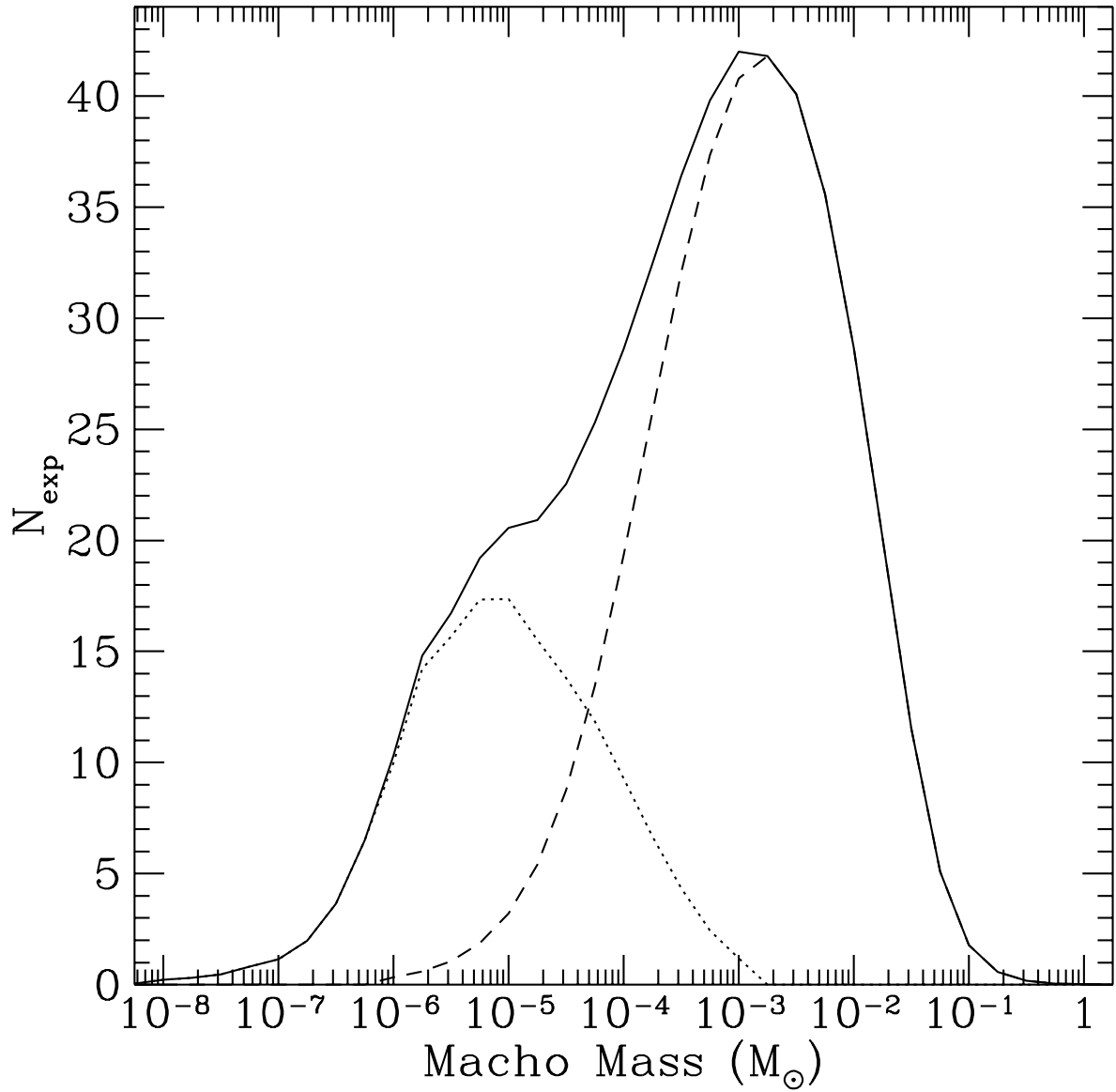}
\caption{Number of expected events as a function of
Macho mass after combining the standard and spike analyses,
with a cut on events with $\that > 20 \,\rm days$.
The
spike result is plotted with the dotted line
(with the double counted events subtracted), the standard result
with the dashed line, and the combined result is shown with the
solid line.}
\label{fignexp_cow}
\end{figure}

\begin{figure}
\plotone{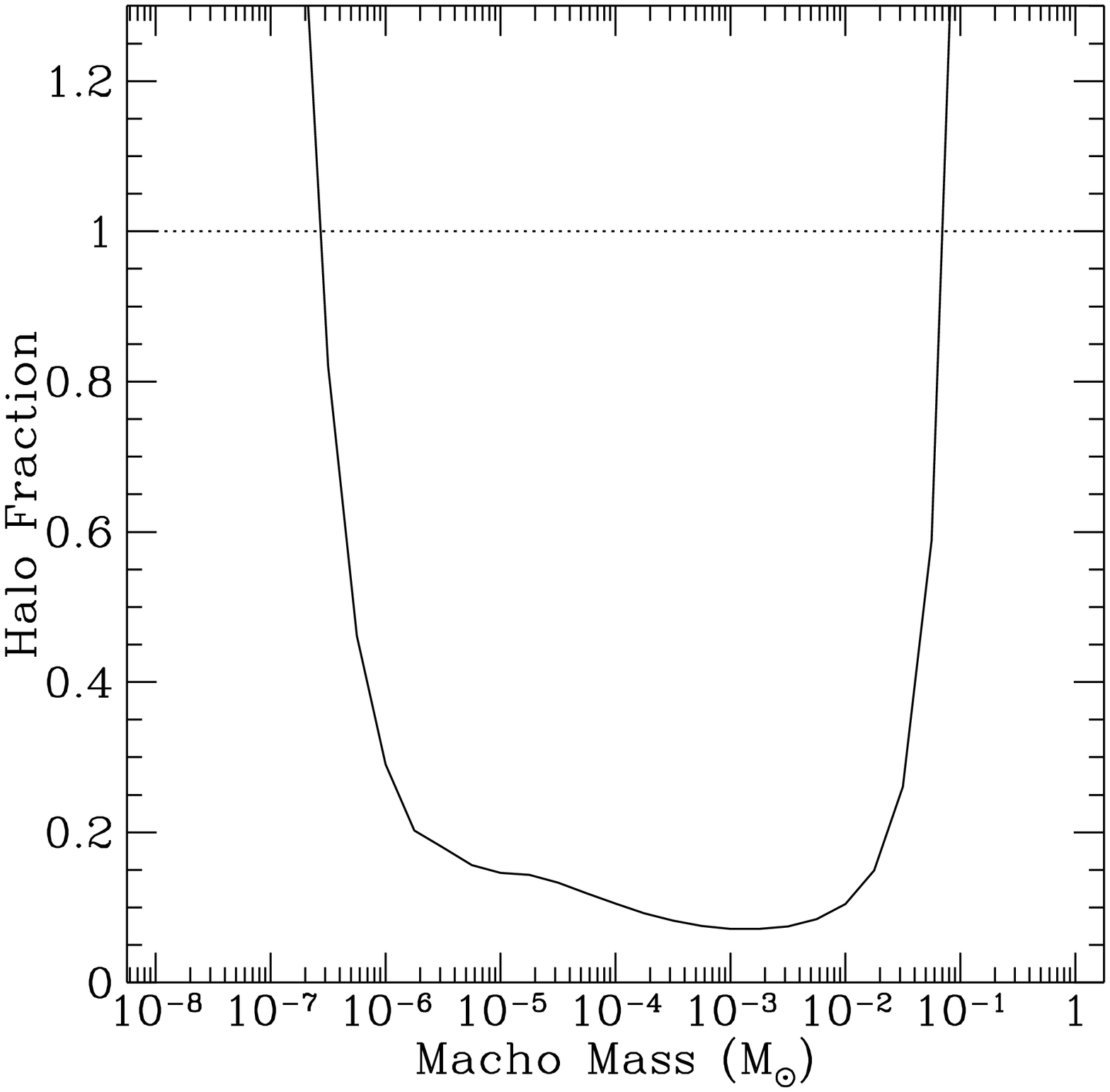}
\caption{Halo fraction upper limit (95\% c.l.) for the
combined spike and standard analyses
and with $\that < 20 \,\rm days$.}
\label{figfrac_cow}
\end{figure}

The standard analysis method of fitting microlensing curves to the data is
sensitive to Machos of masses $10^{-5}\,\msun < m < 1\,\rm \msun$, and it would
be useful to combine the results of the two types of analyses.
To avoid double counting of events which could pass both the
spike and  standard cuts, we ran the standard analysis cuts
on any simulated events passing the spike event cuts,
and any events passing both sets of cuts are so flagged. 
The efficiency is
then recalculated with the flagged events considered
as failing the cuts, and the number of
expected events can then be added to the number of expected events from the 
standard analysis.
However, when adding the number of expected events the number of observed
events must also be added. The standard analysis of the first two years of
LMC data yields eight likely microlensing events
\cite{lmc2}
which would make the
combined limit on halo fraction very weak. However, the eight events all
have durations $\that > 34$ days, and it is very unlikely that these
events were caused by machos with $m < 0.1\,\rm\msun$. Therefore the
standard analysis efficiency was recalculated with a cut such that the event
duration must be shorter than $20$ days. ($20$ days was chosen in order to
be conservative.)
The number of expected events
thus drops significantly at $m > 0.1 \msun$, but we also have no observed
events and are able to place strong limits on lower mass objects.
The number of expected events and halo fraction vs lens mass can be found
in Figures \ref{fignexp_cow} and \ref{figfrac_cow}. Here it can be seen that
Machos of masses $2.5\ee{-7}\,\msun < m < 8.1\ee{-2}\,\rm\msun$
can not make up the
entire halo mass, and lenses in the range
$1.88\ee{-6}\,\msun < m < 2.5\ee{-2}\,\msun$
comprise at most 20\% of the halo dark matter. 

\section{Power Law Halo Models}

The halo models of Evans
\cite{evans94b,evans94a}
allow for rising or falling
rotation curves, flattened halos, and various disk contributions
to the total galactic mass. These models are also called ``power law''
models because at large galactic radii $R$, the circular velocity
$v_{\rm circ} \propto R^{-\beta}$ for some
model parameter $\beta$. The parameters used
to describe the mass and velocity distributions in these models are
as follows:
\begin{itemize}
\item[$\beta$]
At large galactic radii, $\beta = 0$ gives a flat rotation curve,
$\beta < 0$ gives rising curve, and $\beta > 0$ gives falling curve.

\item[$q$]
Halo flattening parameter. $q = 1$ gives spherical halo,
$q = 0.7$ represents ellipticity of E6.

\item[$v_0$]
Normalization velocity.

\item[$R_c$]
Galactic core radius. A large $R_c$ gives a massive disk.

\item[$R_0$]
Radius of solar orbit.
\end{itemize}
The differential event rate $d\Gamma/dxdv_\perp$ can be derived
as a function of these parameters
\cite{macho-explore1},
and it is then
straightforward to calculate the number of expected events using equation
(\ref{eqnexp}). Limits were calculated for the same models used in
\citeN{lmc1},
and the parameters used are found in Table \ref{tablepl}.
Also shown for each model
in Table \ref{tablepl} is the total mass inside $50$ kpc
from the center of the Milky Way, which we call $M_{50}$.
\begin{deluxetable}{ccccccc}
\tablecaption{Power Law Halo Model Parameters\label{tablepl}}
\tablehead{
\colhead{Model} & 
\colhead{$\beta$} & 
\colhead{$q$} & 
\colhead{$v_0\,\rm (km/sec)$} & 
\colhead{$R_c \,\rm(kpc)$} & 
\colhead{$R_0 \,\rm(kpc)$ } &
\colhead{$M_{50} \,\rm(10^{11}\msun)$}}
\startdata
S & - & - & - & 5 & 8.5 & 4.13 \nl
A & 0 & 1 & 200 & 5 & 8.5 & 4.62 \nl
B & -0.2 & 1 & 200 & 5 & 8.5 & 7.34 \nl
C & 0.2 & 1 & 180 & 5 & 8.5 & 2.36 \nl
D & 0 & 0.71 & 200 & 5 & 8.5 & 3.74 \nl
E & 0 & 1 & 90 & 20 & 7.0 & 0.82 \nl
F & 0 & 1 & 150 & 25 & 7.9 & 2.10 \nl
G & 0 & 1 & 180 & 20 & 7.9 & 3.26 \nl
\enddata
\end{deluxetable}

Model S is the simple
standard spherical halo described in Section \ref{theory},
and model A is the power law model equivalent.
Model B has a rising rotation curve and a more massive halo, while model C
has a falling curve and a less massive halo. Model D has a flattened halo,
and models E, F, and G have more massive disks. Model E has an extremely
massive disk and a very light halo, and this model is probably inconsistent
with estimates of the mass of the Milky Way. 

\begin{figure}
\plotone{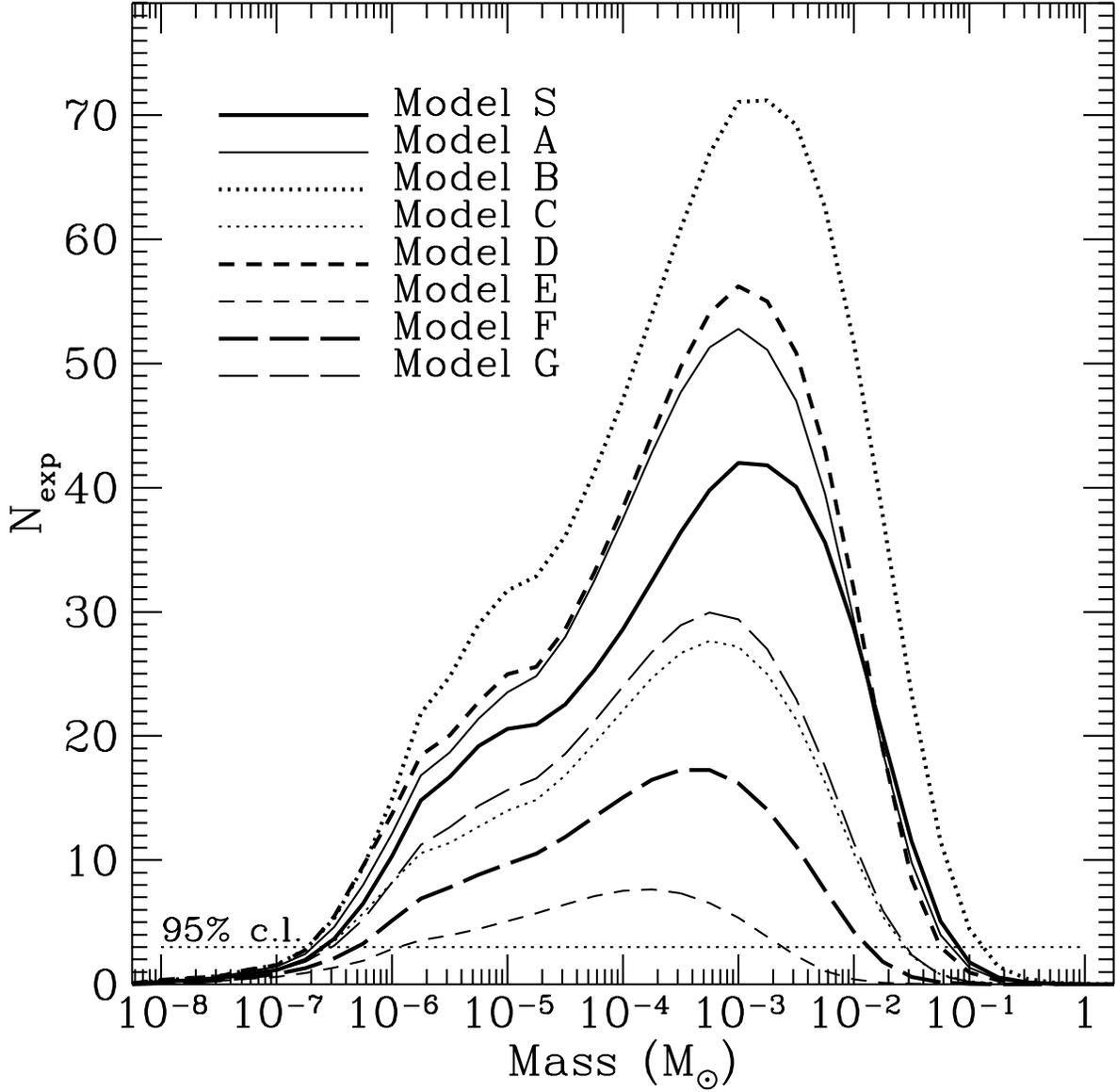}
\caption{A plot of the number of expected events vs
mass for the standard model and seven power law halo models. 
The results shown are for the combined spike and standard analyses,
and with $\that < 20 \,\rm days$.
The
line at $N_{\rm exp} = 3$ is the 95\% c.l. upper limit,
and the regions of the curves above this line are ruled out.}
\label{fignexp_tot}
\end{figure}

\begin{figure}
\plotone{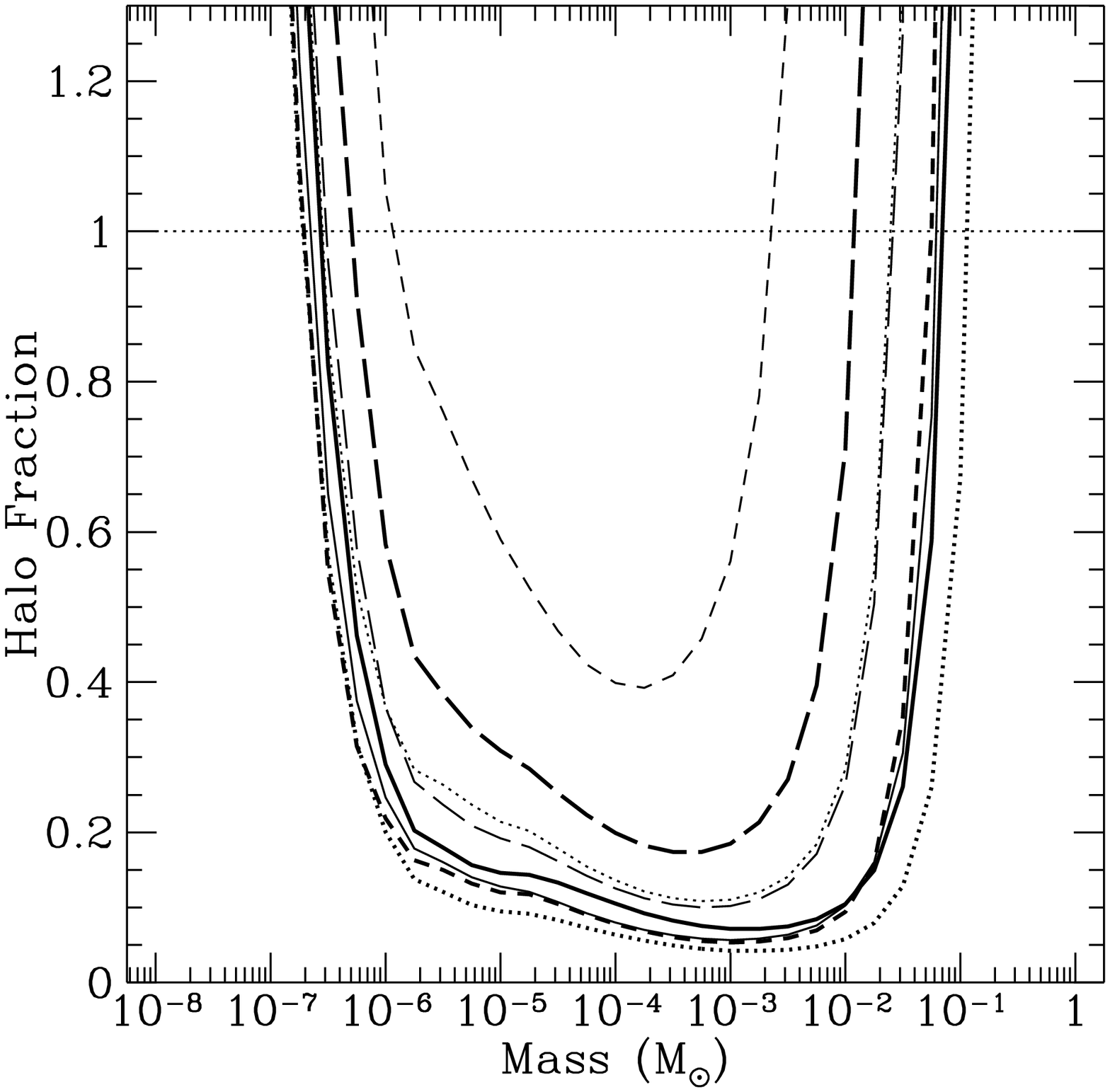}
\caption{Upper limits on Macho fraction of
the halo vs lens mass for the spherical
and power law halo models (line coding is the same as in Figure 10).
The results shown are for the combined spike and standard analyses,
and with $\that < 20 \,\rm days$.
The regions
above the curves are ruled out at the 95\% c.l.}
\label{figfrac_tot}
\end{figure}

The number of expected events was calculated for these models and
then combined with the results from the
standard analysis as described in Section \ref{combanal}.
In Figure \ref{fignexp_tot} we plot the resulting number of expected events
as a function
of lens mass for a $\delta$-function mass distribution using the
simple spherical halo model and the seven power law halo models
described above. Figure \ref{figfrac_tot} is a plot of allowed halo fraction
vs mass for the same models. For the models with more massive halos,
only about 30\% of the halo can be comprised of
Machos in the range of $9.5\ee{-7}\,\msun < m < 2.9\ee{-2} \,\msun$. The
limits get weaker for less massive halos, and little useful parameter
space can be excluded for the extreme ``maximal disk'' model E.

\begin{figure}
\plotone{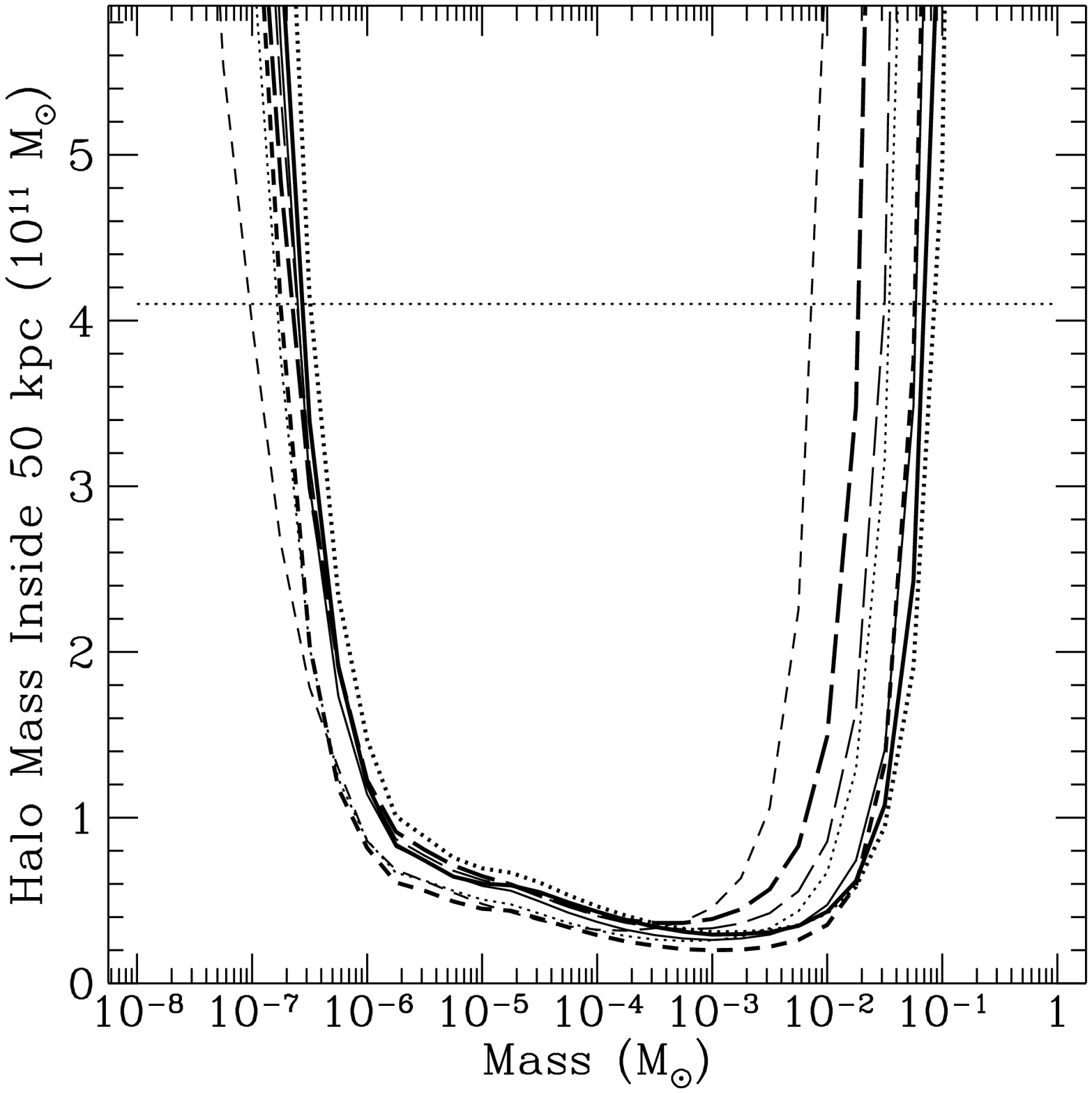}
\caption{Upper limits on the total mass of Machos
interior to $50$ kpc as a function of lens mass.
The results shown are for the combined spike and standard analyses,
and with $\that < 20 \,\rm days$.
The regions above the curves are ruled out at the 95 \% c.l.
Objects of mass $3.2\ee{-7}\,\msun < m < 1.87\ee{-2} \,\msun$
can not make up the
canonical value of $4.1\ee{11} \,\msun$, independent of the model used.}
\label{figmlim_tot}
\end{figure}

The differences between the limits among the various models
is primarily because the number of expected events is directly
proportional to $M_{50}$, or the number of Machos in the halo.
We can get more model-independent limits by removing
this factor and plotting the total allowed halo mass from Machos inside
$50$ kpc, rather than the halo mass fraction, as a function of mass.
This plot is shown in Figure \ref{figmlim_tot}.
Ignoring the unlikely model E we see that, independent of model,
no more than $10^{11} \,\msun$ of the halo mass
inside 50 kpc can come from objects of mass
$1.85\ee{-6}\,\msun < m < 6.5\ee{-3} \,\msun$, and objects in the range
$3.2\ee{-7}\,\msun < m < 1.87\ee{-2} \,\msun$ can not make up the entire 
canonical value of $4.1\ee{11} \,\msun$.

\section{Conclusion}
We have extended the sensitivity of the MACHO experiment
to two orders of magnitude lower in mass using existing data and without
changing observing strategy. Objects with masses
$2.5\ee{-7}\,\msun < m < 8.1\ee{-2}\,\msun$
(roughly one Mars mass to 80 Jupiter
masses) can not comprise the entire standard spherical halo mass,
and Machos in the range $1.88\ee{-6}\,\msun < m < 2.5\ee{-2} \,\msun$ make
up less than 20\% of the halo.
Independent of halo model, objects in the range of
$3.2\ee{-7}\,\msun < m < 1.87\ee{-2}\,\msun$ can not make up
the canonical halo mass 
inside $50$ kpc of $4.1\ee{11}\,\msun$, and less than $10^{11} \,\msun$
of the halo is made from Machos with masses
$1.85\ee{-6}\,\msun < m < 6.5\ee{-3}\,\msun$.
These limits are the strongest published to date.

\acknowledgements
We are grateful for the support given our project by the technical
staff at the Mt. Stromlo Observatory.  Work performed at LLNL is
supported by the DOE under contract W-7405-ENG-48.  Work performed by the
Center for Particle Astrophysics personnel is supported by the NSF
through AST 9120005.  The work at MSSSO is supported by the Australian
Department of Industry, Science, and Technology.  
K.G. acknowledges support from DOE OJI, Alfred P. Sloan, and Cotrell Scholar
awards.  
C.S. acknowledges the generous support of the Packard and Sloan Foundations.
W.S. is supported by a PPARC Advanced fellowship.
M.L. thanks Thor Vandehei for assistance in running the analysis.


\end{document}